\documentclass[fleqn,10pt]{wlscirep}
\usepackage[utf8]{inputenc}
\usepackage[T1]{fontenc}
\usepackage{siunitx}
\title{Bullseye dielectric cavities for photon collection from a surface-mounted quantum-light-emitter}

\author[1]{Reza Hekmati}
\author[2]{John P. Hadden}
\author[1]{Annie Mathew}
\author[2]{Samuel G. Bishop}
\author[1]{Stephen A. Lynch}
\author[1,2,*]{Anthony J. Bennett}
\affil[1]{School of Physics and Astronomy, Cardiff University, Queen’s Buildings, Cardiff CF24 3AA, United Kingdom}
\affil[2]{School of Engineering, Cardiff University, Queen’s Buildings, Cardiff CF24 3AA, United Kingdom}

\affil[*]{BennettA19@cardiff.ac.uk}

\begin{abstract}
Coupling light from a point source to a propagating mode is an important problem in nano-photonics and is essential for many applications in quantum optics. Circular "bullseye" cavities, consisting of concentric rings of alternating refractive index, are a promising technology that can achieve near-unity coupling into a first lens. Here we design a bullseye structure suitable for enhancing the emission from dye molecules, 2D materials and nano-diamonds positioned on the surface of these cavities. A periodic design of cavity, meeting the Bragg scattering condition, achieves a Purcell factor of 22.5 and collection efficiency of 80 $\%$. We also tackle the more challenging task of designing a cavity for coupling to a low numerical aperture fibre in the near field. Using an iterative procedure, we show that apodized (non-periodic) rings can achieve a collection efficiency that exceeds the periodic Bragg cavity.
\end{abstract}
\begin{document}

\flushbottom
\maketitle

\thispagestyle{empty}

\section*{Introduction}
The ability to engineer the emission characteristics of point-like emitters is crucial in the creation of single photon sources for applications in quantum computing and
quantum communication \cite{Barnes2002Solid-stateStrategies, Kuhn2010Cavity-basedSources}. The source's photon collection efficiency and the emitter decay rate are two important figures of merit that can be tailored by cavity design. Numerous cavity designs have been thoroughly investigated such as photonic crystals \cite{Noda, Englund}, micro-pillars \cite{Santori, Bennett} and nano-antenna \cite{Chikkaraddy2016Single-moleculeNanocavities, Sapienza2015NanoscaleEmission}. 

Bullseye cavities, so called because of their resemblance to a bullseye target, consist of concentric rings of alternating dielectrics. These have recently attracted considerable attention due to their unique ability to enhance and direct emission from single photon sources into the far field over a broad spectral range \cite{Andersen2018HybridCollection, Wang2019On-DemandIndistinguishability, Duong2018EnhancedGratings, Zheng2017ChirpedDiamond}. These bullseye-cavities can offer enhancement to the photon collection efficiency and decay rate which is independent of the orientation within the plane of an ideally positioned emitter \cite{Davanco2011AEmission}. To date, reported bullseye structures are mainly dielectric-based and lossless, a key advantage over plasmonic cavities, which include absorptive metals in the high-field region \cite{Chikkaraddy2016Single-moleculeNanocavities, Russell2012LargeNanocavities, Hoang2015UltrafastNanocavities}. These advantages have led bullseyes to be used to enhance the performance of quantum dots \cite{Sapienza2015NanoscaleEmission}, defect centres in diamond \cite{Li2015EfficientGrating,Waltrich2021AAntenna}, and in two dimensional (2D) materials \cite{Duong2018EnhancedGratings, Iff2021Purcell-EnhancedCavity}. Reported values for spontaneous emission rate enhancement in the literature for the bullseye structures are as high as 56 \cite{Zheng2017ChirpedDiamond, Rickert2019OptimizedGratings}.

Recently, a hybrid structure based on a bullseye has been reported in which a high Purcell factor over a broad range of wavelengths with high collection efficiency is achieved \cite{Abudayyeh2020OvercomingSource}. Inspired by the bullseye pattern, defects in photonic crystals were recently proposed as a new platform for extremely high Purcell factor with high collection efficiency \cite{Li2021Field-basedInterfaces}.  A collection efficiency near unity can be achieved by utilising a low refractive-index SiO\textsubscript{2} dielectric layer and a layer of gold as the mirror at the bottom of the structure \cite{Wang2019TowardsMicrocavities}.

In this study, we demonstrate an approach to designing a bullseye structure to enhance the emission from point sources just above the centre of the structure. This case is applicable to emitters in WSe\textsubscript{2} flakes laid over the cavity, single dye molecules or nano-diamonds applied to the structure. The resultant cavity achieeves high collection efficiency ($ >$ 80 $\%$) and high Purcell factor ($ >$ 22.5) at the cavity resonance ( \SI{750}{\nano\metre}). Subsequently, the coupling condition between the emitter and the cavity is discussed where this analysis underlines the importance of positioning in emitter-cavity systems. Finally, we apply the concept of apodization to achieve higher collection efficiency into a low numerical aperture fibre directly aligned to the cavity.

\subsection*{Cavity design}

\begin{figure}[t!]
\centering
\includegraphics[scale=0.28]{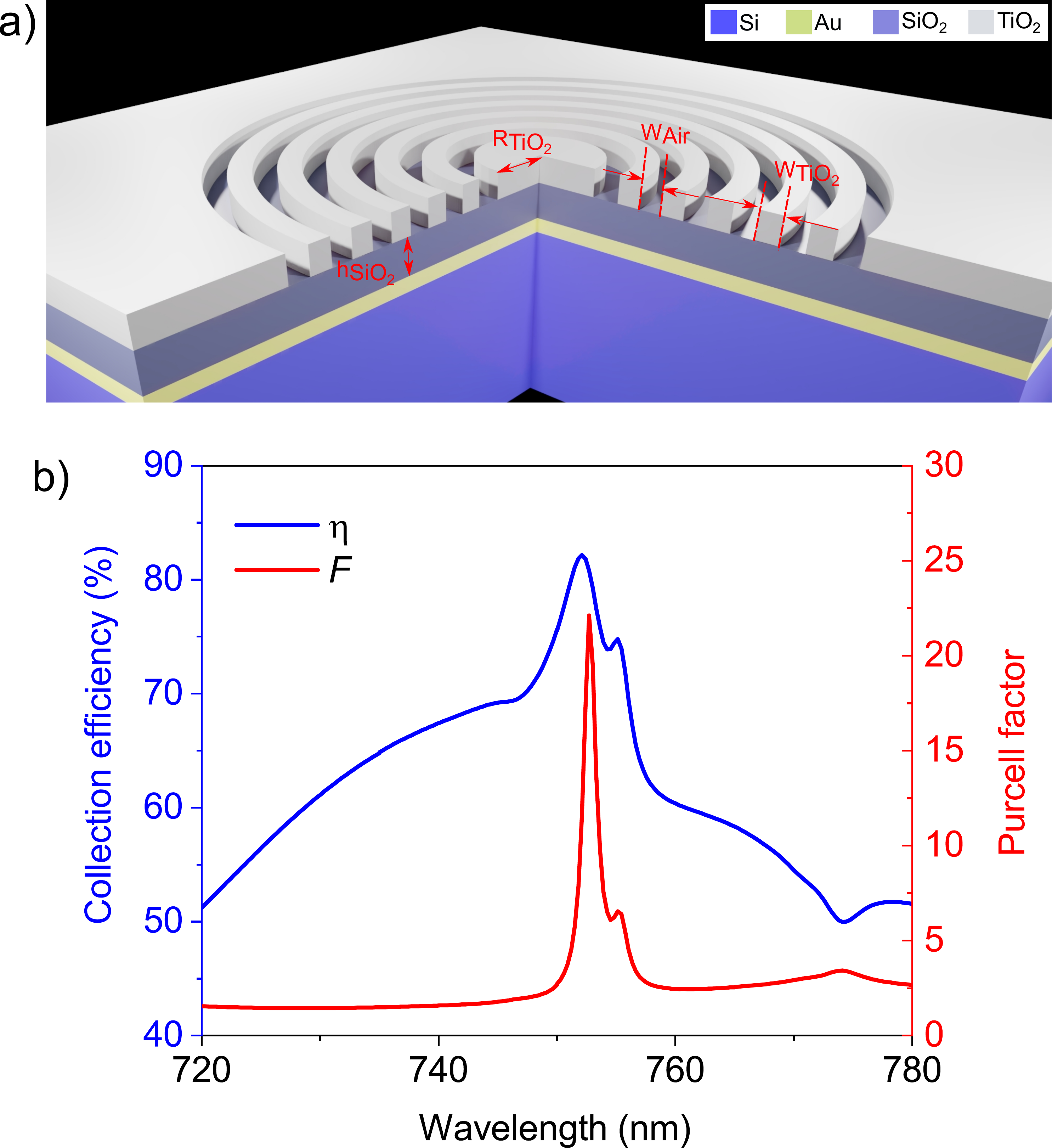}
\caption{Bullseye dielectric cavity structure. (a) 3D representation of the structure, consisting of $TiO\textsubscript{2}$ rings on a $SiO\textsubscript{2}$ layer, above a gold mirror. The key dimensions of the structure are labelled. A dipole emitter is in the middle of the central disc.  (b) Collection efficiency (left y-axis) and the Purcell factor (right y-axis) as a function of the wavelength for a 10 ring-pair cavity designed for  \SI{752} {\nano\metre}.}
\label{Fig:Figure1}
\end{figure}

The structure in this work is depicted in Figure \ref{Fig:Figure1}.a. It consists of a silicon substrate coated in a \SI{150} {\nano\metre} reflective gold layer, $h_{SiO\textsubscript{2}}= \SI{435}{\nano\metre}$ of SiO\textsubscript{2} and $h_{TiO\textsubscript{2}}$ = \SI{200} {\nano\metre} of TiO\textsubscript{2}. A circular grating etched into the TiO\textsubscript{2} layer consisting of a series of concentric rings centred around a disk with a radius of $R_{TiO\textsubscript{2}}$. The width of each ring is $W_{TiO\textsubscript{2}}$ and the distance between two consecutive rings is $W_{Air}$. The emitter is modelled as a dipole in the middle of the TiO\textsubscript{2} at the centre of a bullseye, using the Finite Difference Time Domain package Lumerical. The Au layer acts as a mirror that helps to redirect emission toward the collection optics.

A periodic circular grating, which satisfies the second-order Bragg condition for efficient vertical light extraction, should fulfil the phase matching condition for the periodicity, $\Lambda = W_{TiO\textsubscript{2}} + W_{air}=m\lambda /2 n_{TE}$ \cite{Davanco2011AEmission, Ates2012BrightMicrocavity}. In this formula, $m$, $\lambda$ and $n_{TE}$ are an integer constant ($m=$1), the resonance wavelength and the effective transverse electric (TE) mode index propagating inside the slab, respectively. As explained in Ates $et$ $al$ \cite{Ates2012BrightMicrocavity}, a fully etched grating provides better overlap of the guided field inside the slab and the etched region, higher guided wave reflectivity, and higher quality factor, leading to a higher Purcell factor. Therefore, $\Lambda=\lambda /n_{TE}$ is considered as the initial value for the grating period. Figure \ref{Fig:Figure1}.b. shows the result of a simulation for a cavity with periodic rings as a function of the dipole wavelength, which achieves a maximum Purcell factor of 22.5 and a collection efficiency of 80 $\%$ on resonance. In the following section we discuss in detail the design parameters explored to achieve the aforementioned cavity performance.

\section*{Results}

\subsection*{Simulations of an ideal periodic circular Bragg grating}

The circular Bragg grating exhibits high directionality of the radiated light from an emitter positioned in the centre of the structure. For concreteness, we refer to the collection efficiency, $\eta$, as the fraction of the total dipole power, normalised by the power injected into the simulation, which is collected within a fixed numerical aperture (NA $=$ 0.68), centred on the cavity. 

In dielectric cavities the loss is considered to be negligible; therefore, the spontaneous emission rate enhancement is equal to the Purcell factor \cite{Krasnok2015AnEffect}. In experiments, the Purcell factor quantifies the reduction in the dipole transition's radiative lifetime \cite{Crook2020PurcellControl}. In simulations, the Purcell factor may be be determined by the ratio of the total radiated power of a dipole inside the cavity, divided by the total radiated power of the dipole in an uniform dielectric.

Key parameters in the design of the structure are the grating period, $\Lambda=$  \SI{420} {\nano\metre}, the duty cycle $W_{TiO\textsubscript{2}}/\Lambda$ = 0.67, and the central disk diameter. Figure \ref{Fig:Figure2}.a-d illustrates the Purcell factor and collection efficiency, respectively, as a function of these parameters. We observe the resonances in the Purcell factor are narrower than in the collection efficiency; this means the Purcell factor is highly susceptible to the designed wavelength, highlighting the importance of accurate fabrication (also evident in Figure \ref{Fig:Figure1}b). In the case of the collection efficiency (Figure \ref{Fig:Figure2}.c-d) the gold layer helps ensure some enhancement across a broad range of wavelengths. For a resonance wavelength of  \SI{750} {\nano\metre} the highest performance is achieved with  $R_{TiO\textsubscript{2}}=$ \SI{835} {\nano\metre} and duty cycle 0.66. It is worth noting that the design parameters that correspond to the maximum Purcell effect do not correspond to the maximum collection efficiency. For example, in  Figure \ref{Fig:Figure2}.b and d below a duty cycle of 0.63. This highlights a key point in cavity design: enhanced scattering of light into a lens is a loss mechanism that can reduce the intensity of light trapped in the cavity, and therefore the Purcell effect.

\subsection*{The dipole-cavity coupling condition}

\begin{figure}[b!]
\centering
\includegraphics[scale=0.31]{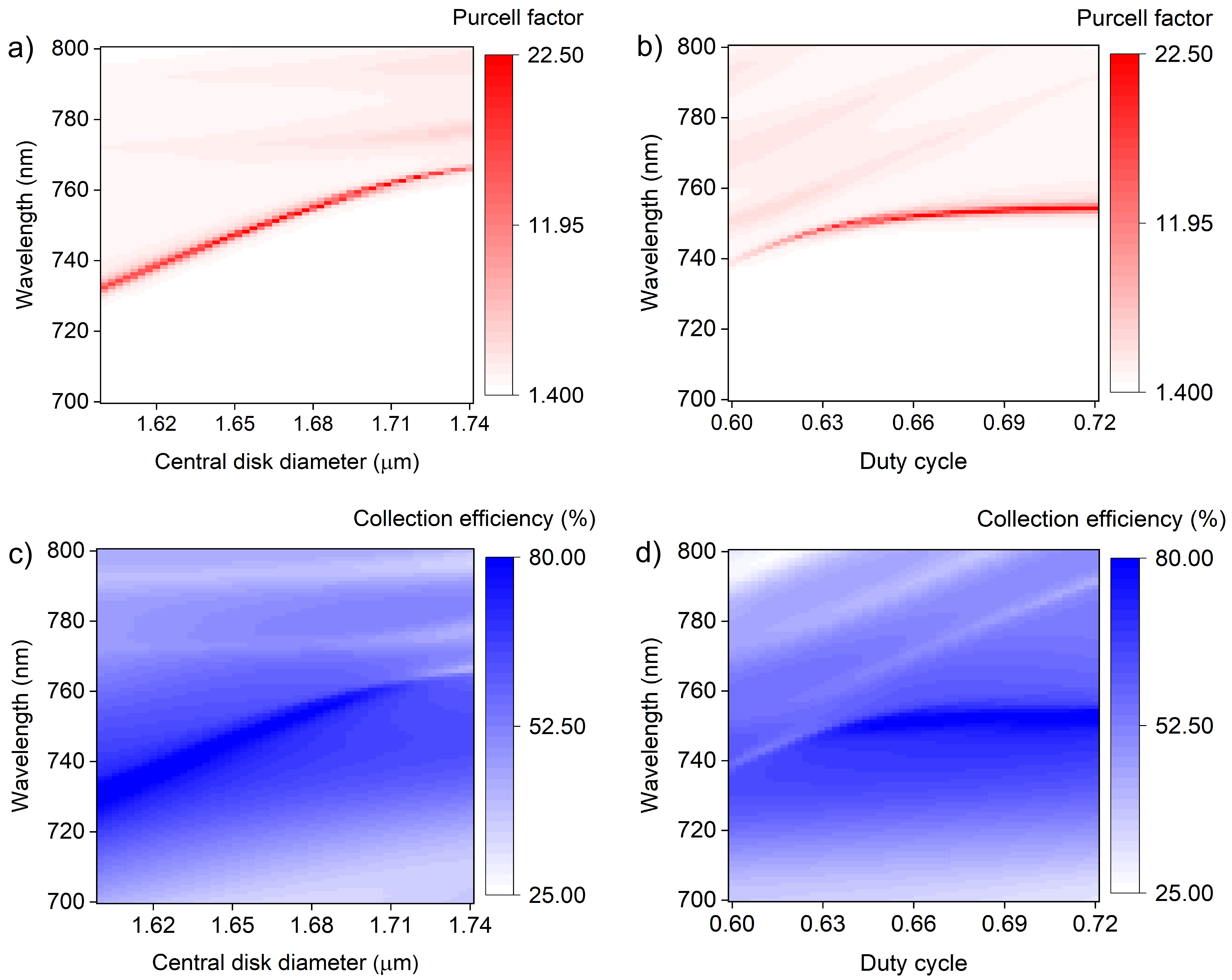}
\caption{Bullseye cavity design. (a,b) Purcell factor as a function of the central disk diameter (a) and the duty cycle (b), and wavelength. (c,d) Collection efficiency into an ideal lens with an numerical aperture of 0.68 as a function of the central disk diameter (c), and the duty cycle (d) and wavelength. The optimal design for a resonance wavelength of \SI{750} {\nano\metre} is achieved with a central disk size of \SI{1.67} {\micro\metre} and a duty cycle of 0.6. The simulated cavity contains 10 rings-periods.}
\label{Fig:Figure2}
\end{figure}

\begin{figure}[t!]
\centering
\includegraphics[scale=0.31]{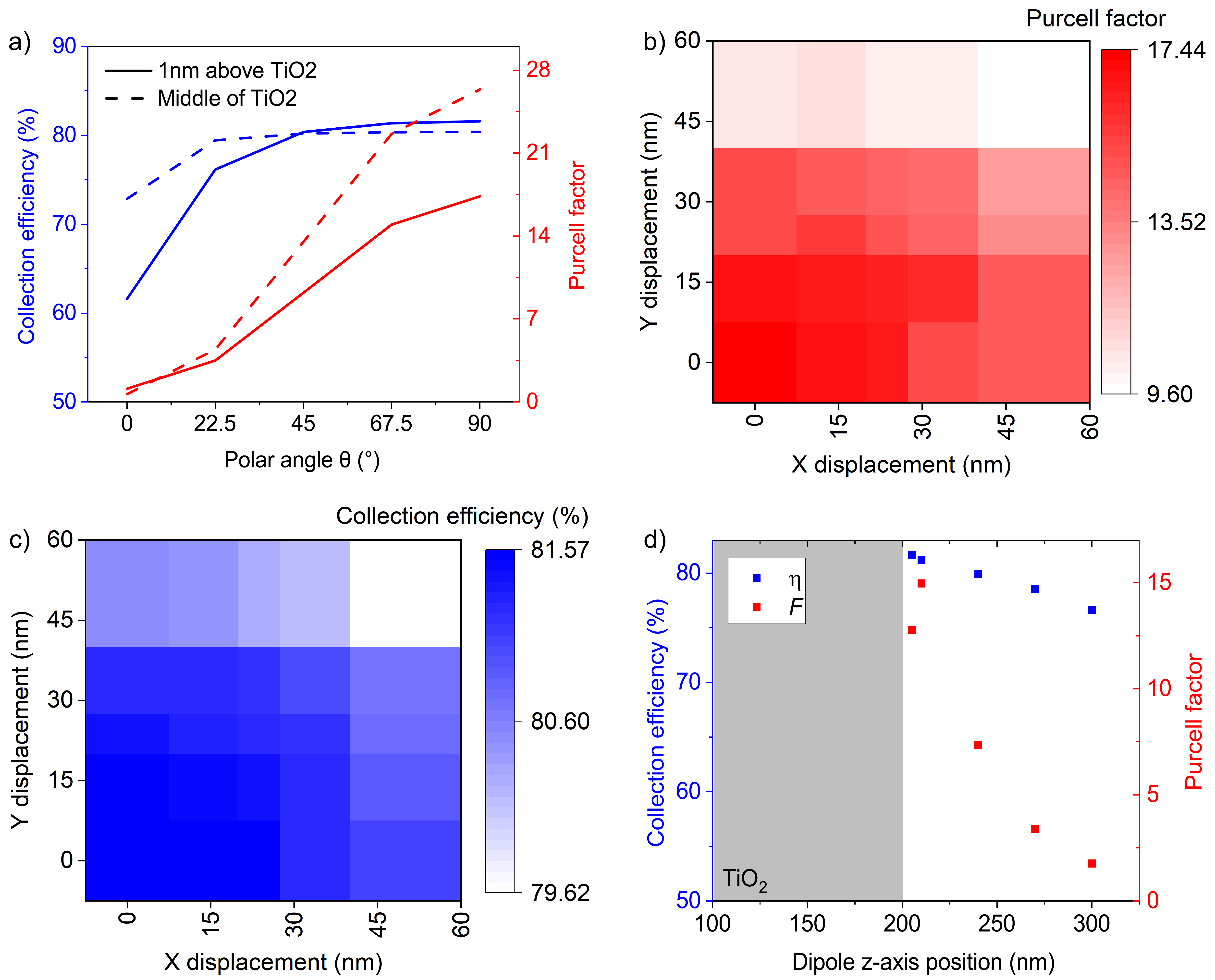}
\caption{The effect of dipole orientation and displacement in the Bullseye cavity. (a) The Purcell factor as a function of the azimuthal angle of the dipole. 0\text{\textdegree} and 90\text{\textdegree} represent out-of-plane and in-plane dipoles, respectively. (b) Purcell factor as a function of the dipole displacement in x and y-direction. (c) Collection efficiency as a function of the dipole displacement in x and y-direction. (d) Collection efficiency (left y-axis) and Purcell factor (right y-axis) as a function of the dipole position along the z-axis (x=y=0). The system resonates at \SI{752} {\nano\metre}. Number of rings in both simulations is 10.}
\label{Fig:Figure3}
\end{figure}

Considering that suitable design parameters have been determined, one can explore the coupling relationship between the dipole and cavity. The formula that explains this relationship is given by $F$, the Purcell factor\cite{Unitt2005PolarizationEffect, Fox2006QuantumIntroduction, Crook2020PurcellControl}:

\begin{equation} \label{Form:Purcell}
F = \cfrac{\tau_{free}}{\tau_{cavity}} = f(\Delta).(\cfrac{\overrightarrow{d}.\overrightarrow{\epsilon}}{|\overrightarrow{d}||\overrightarrow{\epsilon}|})^2 .\cfrac{|\overrightarrow{\epsilon} (\overrightarrow{r})|^2}{|\overrightarrow{\epsilon}_{max}|^2}.F_{P}
\end{equation} 

The first term in Equation \ref{Form:Purcell} describes the spectral detuning between the emitter and the cavity mode, and it is a function of the emitter-field spectral detuning ($\Delta$). The dipole-cavity relation in terms of the wavelength is depicted in Figure \ref{Fig:Figure1}.b. When the dipole and the cavity are wavelength-matched, a Purcell factor of 22.5 and the collection efficiency of 80 $\%$ are achieved. According to the Figure \ref{Fig:Figure2}.a, detuning the dipole wavelength away from the cavity resonance presents, on average, a CE greater than 70 $\%$ over a \SI{4} {\nano\metre} bandwidth, and a CE above 60 $\%$ over a \SI{25.4} {\nano\metre} bandwidth.

The second term in Equation \ref{Form:Purcell} quantifies the alignment between the dipole polarisation ($\overrightarrow{\epsilon}$) and the field in the cavity mode at that position ($\overrightarrow{d}$) \cite{Fox2006QuantumIntroduction}. The optimised structure was designed for an in-plane dipole, $\theta=90\text{\textdegree}$, in which $\theta$ is the azimuthal angle between the z-axis and the dipole. Collection efficiency and the Purcell factor are shown as red and blue in Figure \ref{Fig:Figure3}.a. In the figure, two cases are investigated; when the dipole is placed \SI{1} {\nano\metre} above the TiO\textsubscript{2} layer (solid lines) and when the dipole is in the centre ($z=$\SI{100} {\nano\metre}) of the TiO\textsubscript{2} disk (dashed lines). We see that the higher the angle of $\theta$, the better the value of the collection efficiency and the Purcell factor. However, even for a dipole with vertical orientation the CE is above 60 $\%$ which is 2 order of magnitude greater than in the absence of the cavity.

The third term in Equation \ref{Form:Purcell} describes the spatial detuning of the emitter with respect to the cavity can be explained by the ratio of the electric field ($\epsilon$) at the emitter spatial position $r$, relative to the maximum electric field ($\epsilon_{max}$). Figure \ref{Fig:Figure3}.b,c shows the Purcell factor and the collection efficiency as a function of the displacement in x and y directions, respectively. The Purcell factor drops by half when the dipole is displaced \SI{70} {\nano\metre} from the centre. It should be noted that the x and y displacement plots are not the same because the dipole is oriented along the x axis. The collection efficiency (red) and the relative Purcell factor (blue) are also shown as a function of displacement in the z-direction, for x=y=0, in Figure \ref{Fig:Figure3}.d. As one might expect, the Purcell factor is greatest at the centre of the TiO\textsubscript{2} disk ($z=$\SI{100} {\nano\metre}), but an appreciable enhancement can be seen when the emitter is at the surface.

Finally, $F_{P}$ is a figure of merit representing the maximum enhancement of the spontaneous emission rate for a source with zero spectral detuning, optimal polarization orientation and ideal positioning of the emitter relative to the mode. When all conditions are achieved the maximum coupling between the emitter and the cavity can be expressed by $F_{P}= 3Q(\lambda_{cav}/n)^3/4\pi^2V_{eff}$, where $Q$ and $V_{eff}$ are the quality factor and the effective mode volume of the cavity, respectively.

\subsection*{The effect of increased cavity periods}
\begin{figure}[b!]
\centering
\includegraphics[scale=0.295]{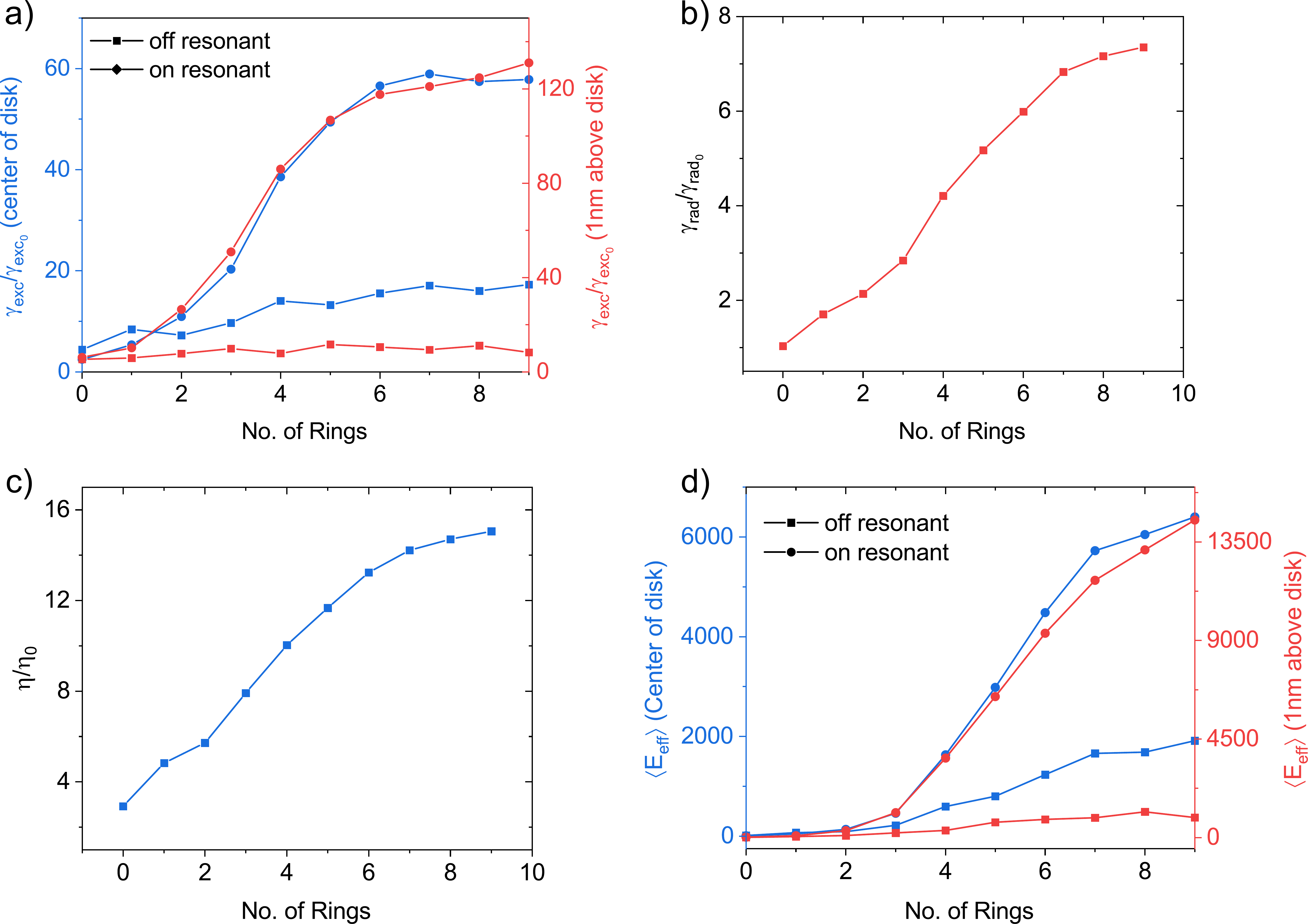}
\caption{Dipole-cavity coupling efficiency. (a) Excitation enhancement rate for a dipole in the centre of the cavity (red) and a dipole \SI{1} {\nano\metre} above the disk (blue) as a function of the number of rings. Square and circular shapes illustrate off-resonance (\SI{532} {\nano\metre}) and on-resonance (\SI{752.5} {\nano\metre}) excitation, respectively. (b) The radiative decay rate enhancement as a function of the number of rings. (c) The collection efficiency enhancement as a function of the number of rings. In all plots, one can observe the general trend of increasing performance of the cavity with more rings. (d) Effective enhancement for a dipole positioned at the centre of disk (blue) and a dipole positioned \SI{1} {\nano\metre} above the disk (red).}
\label{Fig:Figure4}
\end{figure}

Figure \ref{Fig:Figure4}.a illustrates the Purcell factor and collection efficiency change with the number of rings in the bullseye structure. Two excitation wavelengths were investigated; \SI{532} {\nano\metre} (relevant to off-resonance excitation) and \SI{752.5} {\nano\metre} (on-resonance excitation, at the wavelength of the cavity). At both wavelengths, increasing the number of rings leads to a greater enhancement; however, when the dipole is placed \SI{1} {\nano\metre} on top of the disk, resonance excitation enhances by a factor of $\sim$ 13 in comparison to off-resonance excitation. For the dipole embedded at the centre of the disk, this ratio is approximately 4. Figure \ref{Fig:Figure4}.b shows the relative decay rate enhancement as a function of the number of rings. Finally, the collection efficiency ($\eta$) is shown in Figure \ref{Fig:Figure4}.c.

Increasing the number of rings increases these separate figures of merit. The product of these terms results in the enhancement of the photon rate extracted under CW excitation. For on-resonant (off-resonant) excitation, this enhancement tends to a value of 15000 (6900) as the number of rings increases.

\subsection*{Increasing collection efficiency into an optical fibre by apodization}

\begin{figure}[b!]
\centering
\includegraphics[scale=0.3]{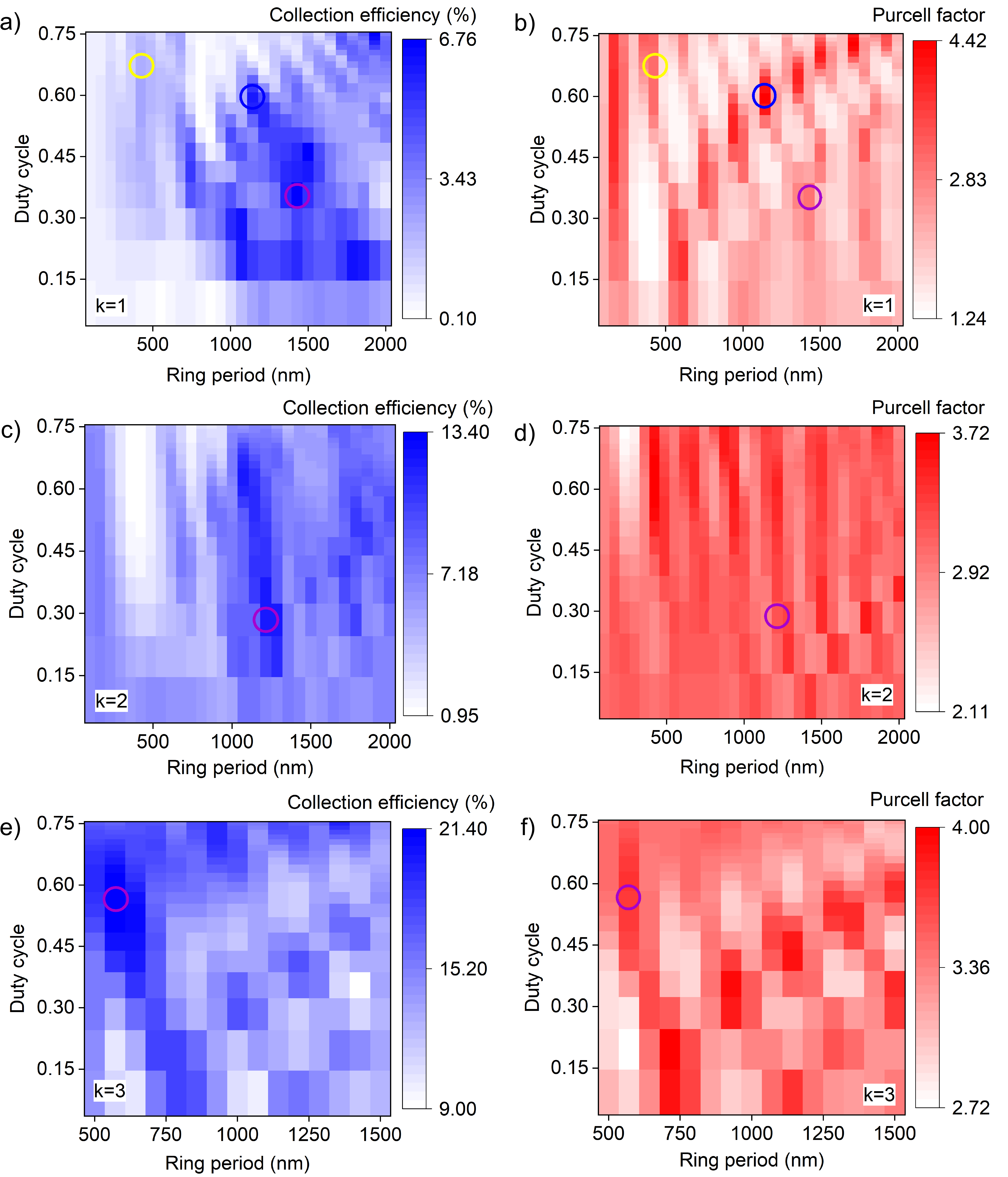}
\caption{Apodization results for the bullseye structure. (a,c,e) Collection efficiency and (b,d,f) Purcell factor as a function of ring period and duty cycle for the first ring (k=1), the second ring (k=2), and the third ring (k=3), respectively. Circles highlight points where the design meets the Bragg condition (yellow), the highest collection efficiency (purple) and the highest Purcell factor (blue). After the collection efficiency is maximised for a ring, it is fixed in dimension before next ring is added and optimised.}
\label{Fig:Figure5}
\end{figure}

The fact that bullseye cavities are flat, rotationally symmetric and highly directional makes them suitable for direct coupling to single mode fibre. Indeed, directly bonding fibres to these efficient structures could be a route to creating robust, compact and connectorised quantum light sources used in real world applications. However, to obtain high collection efficiency the cavity must match the mode field diameter and NA of the fibre.  Here we design a cavity for this application with concentric non-periodic rings, which allows us to vary the mode extent and NA. We refer to this variation of ring dimensions as apodization. The large number of variables in the design makes finding a global maximum efficiency non-trivial. Therefore, we approach this problem by iteratively varying each ring in turn, as discussed below. We consider a commercially available single-mode fibre, Thorlabs 630HP, with a core radius of \SI{1750} {\nano\metre} and core (cladding), refractive indices of 1.46 (1.45) and NA=0.13, fixed above the structure.

Figure \ref{Fig:Figure5}a depicts collection efficiency as a function of ring period and duty cycle for a cavity with a fixed central disk dimension and only one ring (k=1).  Three areas are highlighted with circles; the yellow circle shows the result for a the single ring that meets the Bragg condition, giving collection efficiency of 2.2 $\%$. The purple circle indicates the highest collection efficiency of 6.7$\%$ (ring period  \SI{1428} {\nano\metre}, duty cycle 0.36). The blue circle indicates the highest achieved Purcell factor of 4.4 (ring period \SI{1100} {\nano\metre}, duty cycle 0.60). Again, we notice that the highest collection efficiency does not coincide with the highest Purcell factor, as the additional loss implied by efficient photon collection leads to a lower field in the cavity. We aim to achieve higher collection efficiency, therefore we fix the ring period at \SI{1428} {\nano\metre} and duty cycle 0.36 for the subsequent stages of the iterative design process.

\begin{figure}[t!]
\centering
\includegraphics[scale=0.3]{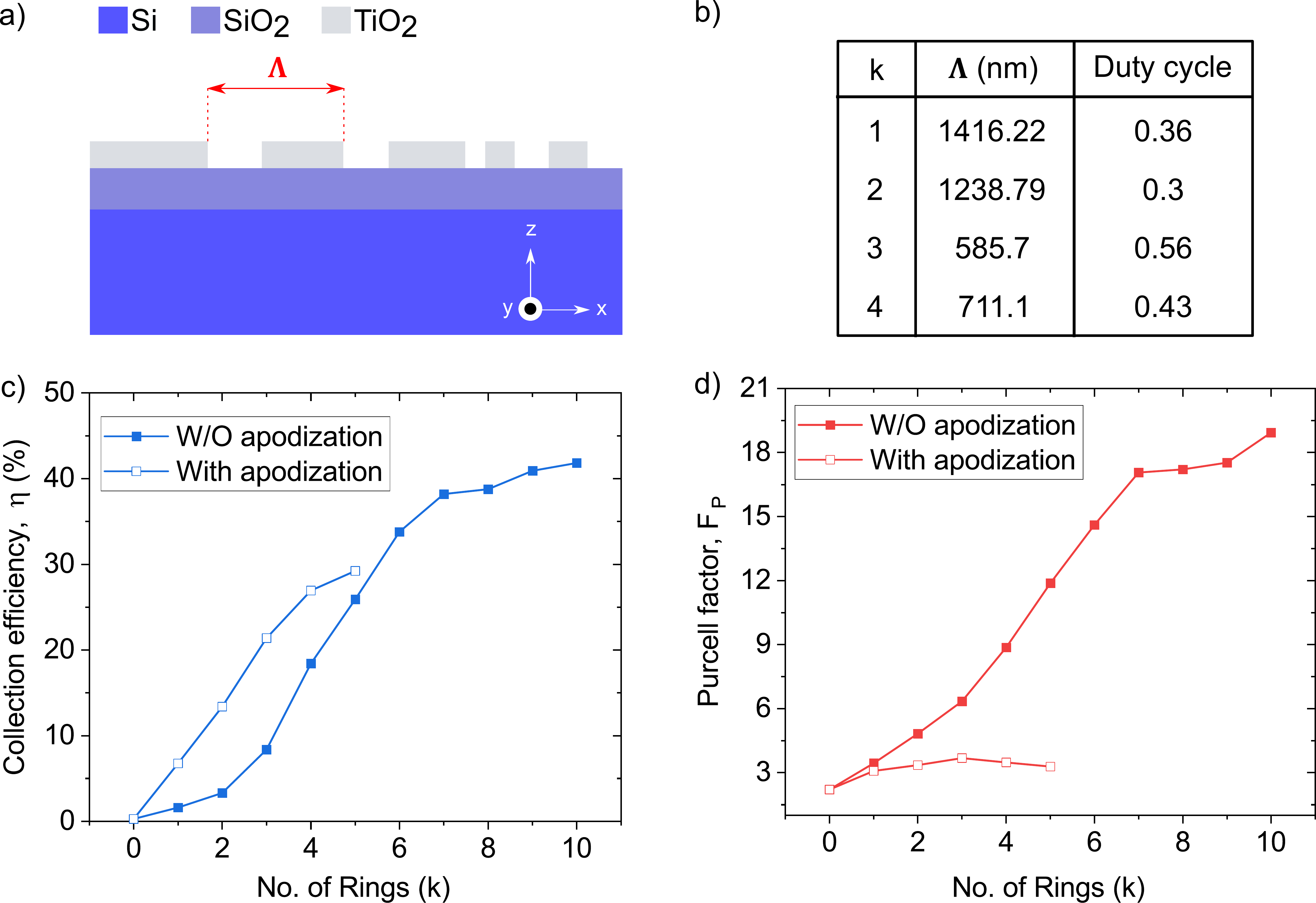}
\caption{Apodised dielectric bullseye cavity coupled to single mode fibre. (a) A cross-section of the apodized bullseye structure. (b) Apodized values for the ring period ($\Lambda$) and duty cycle used in this work. (c) Collection efficiency and (d) Purcell factor as a function of the number of rings (k). In all simulations, the distance between the fibre and the bullseye structure was constant (\SI{2} {\micro\metre}).}
\label{Fig:Figure6}
\end{figure}

We then include a second ring and vary its dimensions, as shown in Figure \ref{Fig:Figure5}.c,d. The blue circle highlights the highest collection efficiency of 13.4 $\%$ for the ring period and duty cycle of \SI{1214} {\nano\metre} and 0.28, respectively. Fixing the dimensions of the second ring, one can then vary the dimensions of the third ring, as shown in Figure \ref{Fig:Figure5}.e,f. We continue with this iterative process up to 5 rings, and the results are shown in Figure \ref{Fig:Figure6}c  compared to the figures of merit for a Bragg cavity with the a same number of rings. It is clear that apodisation leads to high collection efficiencies than that of the uniform Bragg cavity. However, the increasing collection efficiency as a function of the number of rings starts to saturate around k=5. This is to be expected, as at this point the extent of the rings is outside the mode field diameter of the fibre. Figure \ref{Fig:Figure6} a,b summarises the result of this iterative optimisation on the period and duty cycle and Figure \ref{Fig:Figure6}.d shows the resulting Purcell factor for the apodized and non-apodized structures. Underlining the trade-off between achieving high efficiency and high Purcell effect, we see that the efficiency-optimised apodized structure has a lower Purcell effect than the Bragg cavity. It may be possible to further increase the collection efficiency by introducing a snon-uniform etch between rings \cite{Ates2012BrightMicrocavity}.

\section*{Discussion}
We have investigated the ability of "bullseye" cavities to enhance an emitter on the surface of the structure,  which is relevant for future work in nano-optics and quantum technology with 2D materials, dye molecules and nano-diamonds. In a periodic Bragg cavity simulations were used to quantify the accuracy with which the dipole position and the cavity mode must match spatially, spectrally and angularly to achieve high collection efficiency (above 80 $\%$) and Purcell factor (22.5). To optimised the coupling into fibre we considered the design of an apodized cavity, designed using an iterative process. This showed that for structures with less than 5 rings the apodized structure can offer a greater photon collection efficiency than the periodic Bragg cavity. However, the increased scattering into the fibre comes at a cost of reduced field in the cavity, impairing the Purcell factor. We hope that the results of this work will guide future experimental efforts to create efficient, fibre coupled quantum light sources for quantum technology. 


\begin{thebibliography}{10}
\urlstyle{rm}
\expandafter\ifx\csname url\endcsname\relax
  \def\url#1{\texttt{#1}}\fi
\expandafter\ifx\csname urlprefix\endcsname\relax\def\urlprefix{URL }\fi
\expandafter\ifx\csname doiprefix\endcsname\relax\def\doiprefix{DOI: }\fi
\providecommand{\bibinfo}[2]{#2}
\providecommand{\eprint}[2][]{\url{#2}}

\bibitem{Barnes2002Solid-stateStrategies}
\bibinfo{author}{Barnes, W.} \emph{et~al.}
\newblock \bibinfo{journal}{\bibinfo{title}{{Solid-state single photon sources
  : light collection strategies}}}.
\newblock {\emph{\JournalTitle{Eur. Phys. J. D}}}
  \textbf{\bibinfo{volume}{18}}, \bibinfo{pages}{197--210},
  \doiprefix\url{10.1140/epjd/e20020024} (\bibinfo{year}{2002}).

\bibitem{Kuhn2010Cavity-basedSources}
\bibinfo{author}{Kuhn, A.} \& \bibinfo{author}{Ljunggren, D.}
\newblock \bibinfo{journal}{\bibinfo{title}{{Cavity-based single-photon
  sources}}}.
\newblock {\emph{\JournalTitle{Contemporary Physics}}}
  \textbf{\bibinfo{volume}{51}}, \bibinfo{pages}{289--313},
  \doiprefix\url{10.1080/00107511003602990} (\bibinfo{year}{2010}).

  \bibitem{Englund}
\bibinfo{author}{Englund, D.} \emph{et~al.}
\newblock \bibinfo{journal}{\bibinfo{title}{{Controlling the spontaneous emission rate of single quantum dots in a two-dimensional photonic crystal}}}.
\newblock {\emph{\JournalTitle{Nature}}}
  \textbf{\bibinfo{volume}{95}}, \bibinfo{pages}{013904},
  \doiprefix\url{10.1103/PhysRevLett.95.013904} (\bibinfo{year}{2005}).
  
 \bibitem{Noda}
\bibinfo{author}{Akahane, Y.} \emph{et~al.}
\newblock \bibinfo{journal}{\bibinfo{title}{{High-Q photonic nanocavity in a two-dimensional photonic crystal}}}.
\newblock {\emph{\JournalTitle{Physical Review Letters}}}
  \textbf{\bibinfo{volume}{425}}, \bibinfo{pages}{944-947},
  \doiprefix\url{10.1038/nature01086} (\bibinfo{year}{2003}).
  
  \bibitem{Santori}
\bibinfo{author}{Santori, C.} \emph{et~al.}
\newblock \bibinfo{journal}{\bibinfo{title}{{Indistinguishable photons from a single-photon device}}}.
\newblock {\emph{\JournalTitle{Nature Nano.}}}
  \textbf{\bibinfo{volume}{419}}, \bibinfo{pages}{594-597},
  \doiprefix\url{10.1038/nnano.2016.113} (\bibinfo{year}{2002}).
  
\bibitem{Bennett}
\bibinfo{author}{Bennett, A. J.} \emph{et~al.}
\newblock \bibinfo{journal}{\bibinfo{title}{{A semiconductor photon-sorter}}}.
\newblock {\emph{\JournalTitle{Nature Nano.}}}
  \textbf{\bibinfo{volume}{11}}, \bibinfo{pages}{857-860},
  \doiprefix\url{10.1038/nnano.2016.113} (\bibinfo{year}{2016}).
  
\bibitem{Chikkaraddy2016Single-moleculeNanocavities}
\bibinfo{author}{Chikkaraddy, R.} \emph{et~al.}
\newblock \bibinfo{journal}{\bibinfo{title}{{Single-molecule strong coupling at
  room temperature in plasmonic nanocavities}}}.
\newblock {\emph{\JournalTitle{Nature}}} \textbf{\bibinfo{volume}{535}},
  \bibinfo{pages}{127--130}, \doiprefix\url{10.1038/nature17974}
  (\bibinfo{year}{2016}).

\bibitem{Sapienza2015NanoscaleEmission}
\bibinfo{author}{Sapienza, L.}, \bibinfo{author}{Davan{\c{c}}o, M.},
  \bibinfo{author}{Badolato, A.} \& \bibinfo{author}{Srinivasan, K.}
\newblock \bibinfo{journal}{\bibinfo{title}{{Nanoscale optical positioning of
  single quantum dots for bright and pure single-photon emission}}}.
\newblock {\emph{\JournalTitle{Nature Communications}}}
  \textbf{\bibinfo{volume}{6}}, \doiprefix\url{10.1038/ncomms8833}
  (\bibinfo{year}{2015}).
  
\bibitem{Andersen2018HybridCollection}
\bibinfo{author}{Andersen, S.~K.} \emph{et~al.}
\newblock \bibinfo{journal}{\bibinfo{title}{{Hybrid Plasmonic Bullseye Antennas
  for Efficient Photon Collection}}}.
\newblock {\emph{\JournalTitle{ACS Photonics}}} \textbf{\bibinfo{volume}{5}},
  \bibinfo{pages}{692--698}, \doiprefix\url{10.1021/acsphotonics.7b01194}
  (\bibinfo{year}{2018}).

\bibitem{Wang2019On-DemandIndistinguishability}
\bibinfo{author}{Wang, H.} \emph{et~al.}
\newblock \bibinfo{journal}{\bibinfo{title}{{On-demand semiconductor source of entangled photons Which simultaneously has high fidelity, efficiency, and
  indistinguishability}}}.
\newblock {\emph{\JournalTitle{Physical Review Letters}}}
  \textbf{\bibinfo{volume}{122}}, \bibinfo{pages}{1--6},
  \doiprefix\url{10.1103/PhysRevLett.122.113602} (\bibinfo{year}{2019}).

\bibitem{Duong2018EnhancedGratings}
\bibinfo{author}{Duong, N. M.~H.} \emph{et~al.}
\newblock \bibinfo{journal}{\bibinfo{title}{{Enhanced Emission from WSe$_{2}$
  Monolayers Coupled to Circular Bragg Gratings}}}.
\newblock {\emph{\JournalTitle{ACS Photonics}}} \textbf{\bibinfo{volume}{5}},
  \bibinfo{pages}{3950--3955}, \doiprefix\url{10.1021/acsphotonics.8b00865}
  (\bibinfo{year}{2018}).

\bibitem{Zheng2017ChirpedDiamond}
\bibinfo{author}{Zheng, J.}, \bibinfo{author}{Liapis, A.~C.},
  \bibinfo{author}{Chen, E.~H.}, \bibinfo{author}{Black, C.~T.} \&
  \bibinfo{author}{Englund, D.}
\newblock \bibinfo{journal}{\bibinfo{title}{{Chirped circular dielectric
  gratings for near-unity collection efficiency from quantum emitters in bulk
  diamond}}}.
\newblock {\emph{\JournalTitle{Optics Express}}} \textbf{\bibinfo{volume}{89}},
  \bibinfo{pages}{32420--32435},
  \doiprefix\url{10.1103/RevModPhys.89.035002} (\bibinfo{year}{2017}).

\bibitem{Davanco2011AEmission}
\bibinfo{author}{Davan{\c{c}}o, M.}, \bibinfo{author}{Rakher, M.~T.},
  \bibinfo{author}{Schuh, D.}, \bibinfo{author}{Badolato, A.} \&
  \bibinfo{author}{Srinivasan, K.}
\newblock \bibinfo{journal}{\bibinfo{title}{{A circular dielectric grating for
  vertical extraction of single quantum dot emission}}}.
\newblock {\emph{\JournalTitle{Applied Physics Letters}}}
  \textbf{\bibinfo{volume}{99}}, \bibinfo{pages}{4--6},
  \doiprefix\url{10.1063/1.3615051} (\bibinfo{year}{2011}).

\bibitem{Russell2012LargeNanocavities}
\bibinfo{author}{Russell, K.~J.}, \bibinfo{author}{Liu, T.~L.},
  \bibinfo{author}{Cui, S.} \& \bibinfo{author}{Hu, E.~L.}
\newblock \bibinfo{journal}{\bibinfo{title}{{Large spontaneous emission
  enhancement in plasmonic nanocavities}}}.
\newblock {\emph{\JournalTitle{Nature Photonics}}}
  \textbf{\bibinfo{volume}{6}}, \bibinfo{pages}{459--462},
  \doiprefix\url{10.1038/nphoton.2012.112} (\bibinfo{year}{2012}).

\bibitem{Hoang2015UltrafastNanocavities}
\bibinfo{author}{Hoang, T.~B.}, \bibinfo{author}{Akselrod, G.~M.} \&
  \bibinfo{author}{Mikkelsen, M.~H.}
\newblock \bibinfo{journal}{\bibinfo{title}{{Ultrafast Room-Temperature Single
  Photon Emission from Quantum Dots Coupled to Plasmonic Nanocavities}}}.
\newblock {\emph{\JournalTitle{Nano Letters}}} \textbf{\bibinfo{volume}{16}},
  \bibinfo{pages}{270--275}, \doiprefix\url{10.1021/acs.nanolett.5b03724}
  (\bibinfo{year}{2015}).

\bibitem{Abudayyeh2020OvercomingSource}
\bibinfo{author}{Abudayyeh, H.} \emph{et~al.}
\newblock \bibinfo{journal}{\bibinfo{title}{{Overcoming the rate-directionality
  tradeoff: a room-temperature ultrabright quantum light source}}}.
\newblock {\emph{\JournalTitle{ACS Photonics}}} \textbf{\bibinfo{volume}{15}},
  \bibinfo{pages}{17384--17391}, \doiprefix\url{10.1021/acsnano.1c08591}
  (\bibinfo{year}{2021}).

\bibitem{Li2021Field-basedInterfaces}
\bibinfo{author}{Li, L.}, \bibinfo{author}{Choi, H.}, \bibinfo{author}{Heuck,
  M.} \& \bibinfo{author}{Englund, D.}
\newblock \bibinfo{journal}{\bibinfo{title}{{Field-based design of a resonant
  dielectric antenna for coherent spin-photon interfaces}}}.
\newblock {\emph{\JournalTitle{Optics Express}}} \textbf{\bibinfo{volume}{29}},
  \bibinfo{pages}{16469-16476}, \doiprefix\url{10.1364/oe.419773}
  (\bibinfo{year}{2021}).

\bibitem{Li2015EfficientGrating}
\bibinfo{author}{Li, L.} \emph{et~al.}
\newblock \bibinfo{journal}{\bibinfo{title}{{Efficient photon collection from a
  nitrogen vacancy center in a circular bullseye grating}}}.
\newblock {\emph{\JournalTitle{Nano Letters}}}, \textbf{\bibinfo{volume}{15}},
  \bibinfo{pages}{1493--1497}, \doiprefix\url{10.1021/nl503451j}
  (\bibinfo{year}{2015}).

\bibitem{Waltrich2021AAntenna}
\bibinfo{author}{Waltrich, R.} \emph{et~al.}
\newblock \bibinfo{journal}{\bibinfo{title}{{High-purity single photons obtained with moderate-NA optics from SiV center in nanodiamonds on a bullseye antenna}}}.
\newblock {\emph{\JournalTitle{New J. Phys.}}}, \textbf{\bibinfo{volume}{23}},
  \bibinfo{pages}{113022},
  \doiprefix\url{10.1088/1367-2630/ac33f3} (\bibinfo{year}{2021}).

\bibitem{Iff2021Purcell-EnhancedCavity}
\bibinfo{author}{Iff, O.} \emph{et~al.}
\newblock \bibinfo{journal}{\bibinfo{title}{{Purcell-Enhanced Single Photon
  Source Based on a Deterministically Placed WSe2Monolayer Quantum Dot in a
  Circular Bragg Grating Cavity}}}.
\newblock {\emph{\JournalTitle{Nano Letters}}}, \textbf{\bibinfo{volume}{21}},
  \bibinfo{pages}{4715-4720}
  \doiprefix\url{10.1021/acs.nanolett.1c00978} (\bibinfo{year}{2021}).

\bibitem{Wang2019TowardsMicrocavities}
\bibinfo{author}{Wang, H.} \emph{et~al.}
\newblock \bibinfo{journal}{\bibinfo{title}{{Towards optimal single-photon
  sources from polarized microcavities}}}.
\newblock {\emph{\JournalTitle{Nature Photonics}}}
  \textbf{\bibinfo{volume}{13}}, \bibinfo{pages}{770--775},
  \doiprefix\url{10.1038/s41566-019-0494-3} (\bibinfo{year}{2019}).

\bibitem{Ates2012BrightMicrocavity}
\bibinfo{author}{Ates, S.}, \bibinfo{author}{Sapienza, L.},
  \bibinfo{author}{Davanco, M.}, \bibinfo{author}{Badolato, A.} \&
  \bibinfo{author}{Srinivasan, K.}
\newblock \bibinfo{journal}{\bibinfo{title}{{Bright single-photon emission from
  a quantum dot in a circular bragg grating microcavity}}}.
\newblock {\emph{\JournalTitle{IEEE Journal on Selected Topics in Quantum
  Electronics}}} \textbf{\bibinfo{volume}{18}}, \bibinfo{pages}{1711--1721},
  \doiprefix\url{10.1109/JSTQE.2012.2193877} (\bibinfo{year}{2012}).

\bibitem{Rickert2019OptimizedGratings}
\bibinfo{author}{Rickert, L.}, \bibinfo{author}{Kupko, T.},
  \bibinfo{author}{Rodt, S.}, \bibinfo{author}{Reitzenstein, S.} \&
  \bibinfo{author}{Heindel, T.}
\newblock \bibinfo{journal}{\bibinfo{title}{{Optimized designs for
  telecom-wavelength quantum light sources based on hybrid circular Bragg
  gratings}}}.
\newblock {\emph{\JournalTitle{Optics Express}}} \textbf{\bibinfo{volume}{27}},
  \bibinfo{pages}{36824--36837}, \doiprefix\url{10.1364/oe.27.036824}
  (\bibinfo{year}{2019}).

\bibitem{Krasnok2015AnEffect}
\bibinfo{author}{Krasnok, A.~E.} \emph{et~al.}
\newblock \bibinfo{journal}{\bibinfo{title}{{An antenna model for the Purcell
  effect}}}.
\newblock {\emph{\JournalTitle{Scientific Reports}}}
  \textbf{\bibinfo{volume}{5}}, \bibinfo{pages}{1--16},
  \doiprefix\url{10.1038/srep12956} (\bibinfo{year}{2015}).

\bibitem{Crook2020PurcellControl}
\bibinfo{author}{Crook, A.~L.} \emph{et~al.}
\newblock \bibinfo{journal}{\bibinfo{title}{{Purcell enhancement of a single
  silicon carbide color center with coherent spin control}}}.
\newblock {\emph{\JournalTitle{Nano Letters}}} \textbf{\bibinfo{volume}{20}},
  \bibinfo{pages}{3427--3434}, \doiprefix\url{10.1021/acs.nanolett.0c00339}
  (\bibinfo{year}{2020}).

\bibitem{Unitt2005PolarizationEffect}
\bibinfo{author}{Unitt, D.~C.}, \bibinfo{author}{Bennett, A.~J.},
  \bibinfo{author}{Atkinson, P.}, \bibinfo{author}{Ritchie, D.~A.} \&
  \bibinfo{author}{Shields, A.~J.}
\newblock \bibinfo{journal}{\bibinfo{title}{{Polarization control of quantum
  dot single-photon sources via a dipole-dependent Purcell effect}}}.
\newblock {\emph{\JournalTitle{Physical Review B}}} \textbf{\bibinfo{volume}{72}}, \bibinfo{pages}{2--5},
  \doiprefix\url{10.1103/PhysRevB.72.033318} (\bibinfo{year}{2005}).

\bibitem{Fox2006QuantumIntroduction}
\bibinfo{author}{Fox, M.}
\newblock \emph{\bibinfo{title}{{Quantum Optics - An Introduction}}}
  (\bibinfo{publisher}{Oxford University Press}, \bibinfo{year}{2006}).

\bibitem{Gerard2000StrongMicrodisks}
\bibinfo{author}{Gerard, J.-M.} \& \bibinfo{author}{Gayral, B.}
\newblock \bibinfo{journal}{\bibinfo{title}{{Strong Purcell effect for InAs
  quantum boxes in high-Q wet-etched microdisks}}}.
\newblock {\emph{\JournalTitle{Physica E: Low-Dimensional Systems and
  Nanostructures}}} \textbf{\bibinfo{volume}{7}}, \bibinfo{pages}{641--645},
  \doiprefix\url{10.1016/S1386-9477(99)00400-2} (\bibinfo{year}{2000}).

\bibitem{Sortino2019EnhancedNano-antennas}
\bibinfo{author}{Sortino, L.} \emph{et~al.}
\newblock \bibinfo{journal}{\bibinfo{title}{{Enhanced light-matter interaction
  in an atomically thin semiconductor coupled with dielectric nano-antennas}}}.
\newblock {\emph{\JournalTitle{Nature Communications}}}
  \textbf{\bibinfo{volume}{10}}, \doiprefix\url{10.1038/s41467-019-12963-3}
  (\bibinfo{year}{2019}).

\bibitem{Amani2015Near-unityMoS2}
\bibinfo{author}{Amani, M.} \emph{et~al.}
\newblock \bibinfo{journal}{\bibinfo{title}{{Near-unity photoluminescence
  quantum yield in MoS2}}}.
\newblock {\emph{\JournalTitle{Science}}} \textbf{\bibinfo{volume}{350}},
  \bibinfo{pages}{1065--1068}, \doiprefix\url{10.1126/science.aad2114}
  (\bibinfo{year}{2015}).

\bibitem{Koenderink2017Single-PhotonNanoantennas}
\bibinfo{author}{Koenderink, A.~F.}
\newblock \bibinfo{journal}{\bibinfo{title}{{Single-Photon Nanoantennas}}}.
\newblock {\emph{\JournalTitle{ACS Photonics}}} \textbf{\bibinfo{volume}{4}},
  \bibinfo{pages}{710--722}, \doiprefix\url{10.1021/acsphotonics.7b00061}
  (\bibinfo{year}{2017}).

\bibitem{Liu2019AIndistinguishability}
\bibinfo{author}{Liu, J.} \emph{et~al.}
\newblock \bibinfo{journal}{\bibinfo{title}{{A solid-state source of strongly
  entangled photon pairs with high brightness and indistinguishability}}}.
\newblock {\emph{\JournalTitle{Nature Nanotechnology}}}
  \textbf{\bibinfo{volume}{14}}, \bibinfo{pages}{586--593},
  \doiprefix\url{10.1038/s41565-019-0435-9} (\bibinfo{year}{2019}).

\bibitem{Moczaa-Dusanowska2020Strain-TunableDots}
\bibinfo{author}{Moczaa-Dusanowska, M.} \emph{et~al.}
\newblock \bibinfo{journal}{\bibinfo{title}{{Strain-Tunable Single-Photon
  Source Based on a Circular Bragg Grating Cavity with Embedded Quantum
  Dots}}}.
\newblock {\emph{\JournalTitle{ACS Photonics}}} \textbf{\bibinfo{volume}{7}},
  \bibinfo{pages}{3474--3480},
  \doiprefix\url{10.1021/acsphotonics.0c01465} (\bibinfo{year}{2020}).

\bibitem{Barbiero2022DesignGrating}
\bibinfo{author}{Barbiero, A.} \emph{et~al.}
\newblock \bibinfo{journal}{\bibinfo{title}{{Design study for an efficient
  semiconductor quantum light source operating in the telecom C-band based on
  an electrically-driven circular Bragg grating}}}.
\newblock {\emph{\JournalTitle{Optics Express}}} \textbf{\bibinfo{volume}{30}},
  \bibinfo{pages}{10919--10928}, \doiprefix\url{10.1364/oe.452328}
  (\bibinfo{year}{2022}).

\bibitem{Adachi1998OpticalIn1xGaxAsyP1-y}
\bibinfo{author}{Adachi, S.}
\newblock \bibinfo{journal}{\bibinfo{title}{{Optical dispersion relations for
  GaP, GaAs, GaSb, InP, InAs, InSb, AlGaAs, and InGaAs}}}.
\newblock {\emph{\JournalTitle{Journal of Applied Physics}}} \textbf{\bibinfo{volume}{66}},
  \bibinfo{pages}{6030--6040},
  \doiprefix\url{10.1063/1.343580} (\bibinfo{year}{1998}).

\end{thebibliography}

\end{document}